\documentclass[draft]{aipproc}
\layoutstyle{8x11double}

\begin{document}

\title{A 610-MHz Galactic Plane Pulsar Search with the Giant Meterwave
Radio Telescope}


\author{B. C. Joshi}{
  address={National Center for Radio Astrophysics, TIFR, India}
}

\author{M. A. McLaughlin}{
  address={West Virginia University, USA}
}

\author{M. Kramer}{
  address={University of Manchester, Jodrell Bank Observatory, UK}
}

\author{A. G. Lyne}{
  address={University of Manchester, Jodrell Bank Observatory, UK}
}

\author{D. R. Lorimer}{
  address={West Virginia University, USA}
}

\author{D. A. Ludovici}{
  address={West Virginia University, USA}
}

\author{M. Davies}{
  address={Cambridge University, UK}
}

\author{A. J. Faulkner}{
  address={University of Manchester, Jodrell Bank Observatory, UK}
}

\begin{abstract}
We report on the discovery of three new pulsars in 
the first blind survey of the north Galactic plane 
(45$^{\circ}$ < l  < 135$^{\circ}$ ; |b| < 1$^{\circ}$) 
with the Giant Meterwave Radio telescope (GMRT)  at 
an intermediate frequency of  610 MHz. 
The timing parameters, obtained in follow up observations 
with the Lovell Telescope at Jodrell Bank Observatory and 
the GMRT, are  presented. 
\end{abstract}

\maketitle

\section{Introduction}

Most pulsar surveys have been carried out with 
single dish telescopes where there is a trade-off 
between  the collecting area and the beam-width, and 
consequently the rate of the survey. In a 
multi-element telescope such as the Giant Meterwave Radio Telescope (GMRT), 
a large number of smaller antennas can be combined to provide a high sensitivity 
and yet retain a relatively large beam-width. 
In this paper, we report on the discovery of three new pulsars in 
the first blind survey of the north Galactic plane 
(45$^\circ$ < l  < 135$^\circ$ ; |b| < 1$^\circ$) with the GMRT
at an intermediate frequency of  610 MHz, which represents 
the best trade-off between the increased flux density at 
low frequency for pulsars, interstellar scattering and dispersion 
and beam-width. The GMRT's multi-element nature was also exploited to 
determine the positions of the pulsars to an accuracy of 5 arcminutes 
and this technique is also described.

\section{Observations}

The survey consists of 300 fields. 
The observations were conducted using typically 20 to 25 45-m GMRT antennas 
combined in an incoherent array mode at a frequency of 610 MHz.
Each 43' by 43' field in this mode was observed for 35 minutes 
with a bandwidth of 16 MHz with 256 spectral 
channel across the band. 
The data in each channel  were acquired with 16-bit precision every 256 
  $\mu$s after summing the two polarizations and recorded to SDLT tapes 
for off-line processing. 
The 8 sigma threshold for detecting a pulsed signal with a duty cycle of 
10 percent for the configuration used is 0.5 mJy, which is comparable 
to the sensitivity of the Parkes multibeam survey (Manchester et al. 
2001).

\begin{figure}
   \resizebox{0.7\textwidth}{!}
                {\includegraphics{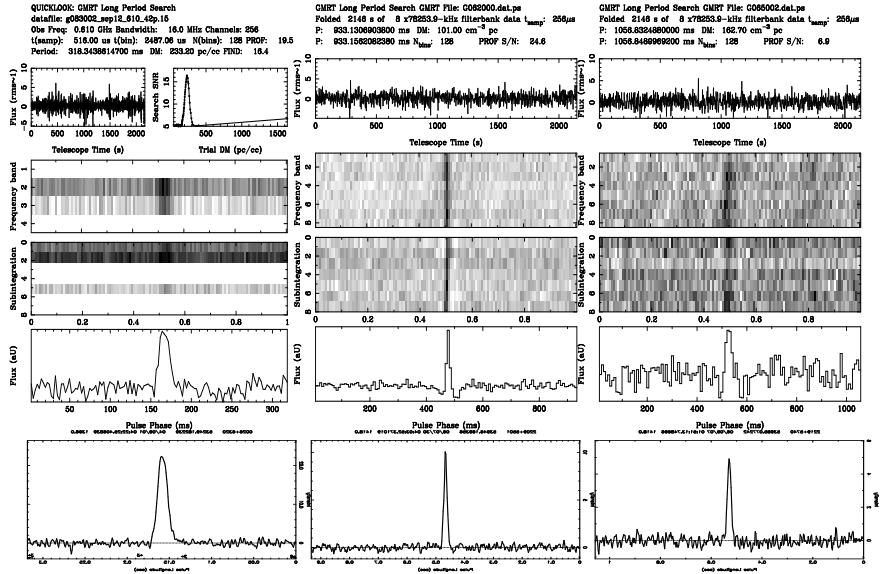}}
    \caption{Discovery plots for the three new pulsars, PSRs J0026+6320, 
         J2208+5500 and J2218+5729 (left to right). The top plot in 
         each panel shows the root mean square power as a function of time
       . The second and third 
         plot in each panel show intensity as a function of subband and 
         pulse phase and sub-integration and pulse phase respectively. 
         The bottom two plots show average profile at 610 MHz observed 
         with the GMRT and at 1420 MHz observed with the Lovell 
         telescope respectively.} 
\end{figure}

\section{Candidate Localization}

The pulsar candidates were confirmed in the follow-up observations 
with the GMRT using the same observing configuration as used for the 
survey. The pulsar position was localized exploiting the multi-element 
nature of the GMRT. The range of baselines available for the GMRT 
antennas allows forming beams with a range of beam-widths 
when appropriate antennas are combined as a phased array 
to form an equivalent single dish with similar sensitivity 
as a 20 antenna incoherent array. Three combinations of the nearest 
3, 5 and 6 antennas respectively were used in this mode to 
observe the candidate field and four fields offset in Right 
Ascension and Declination by half of Full Width at Half 
Maximum (FWHM) for the respective array. 
The respective FWHM  in the above 
configurations were 20, 10 and 5 arcminutes. 
The detected signal-to-noise ratio 
of each new pulsar in these gridding observations 
was used to refine the position successively 
to 5 arcminutes accuracy. The refined position was used for timing  
observations at 1420 MHz with the Lovell Telescope at 
Jodrell Bank Observatory. 
This allowed follow-up confirmation and timing studies with 
high sensitivity with the Lovell Telescope and a rapid 
determination of pulsar parameters.

\begin{table}
  \resizebox{0.8\textwidth}{!}
{\begin{tabular}{l|c|c|c|c|c|c|c}
\hline
NAME &  l$_{field}$ & b$_{field}$ & l  &  b  &  DM  &  P     &  Pdot \\
     & (deg)      & (deg)   &(deg)&(deg)&(pc/cm$^{3}$)&(s)&(10$^{-15}$) \\ \hline
J0026+6320&120.15&0.78&120.18&0.59&230.31&0.318357728337(2)&0.1500(2) \\
J2208+5500&101.25&-0.78&100.94&-0.75&101.03&0.93316093521(1)&6.988(5) \\
J2218+5729&103.95&0.78&103.52&0.49&162.75&1.056844&                  \\ \hline
\end{tabular}}
\caption{Timing Parameters of the new pulsars}
\end{table}

\section{Analysis}

The data were analyzed using the pulsar searching package
SIGPROC   (\url{http://sigproc.sourceforge.net}). The data were
dedispersed using 145 trial dispersion measures (DM) ranging from 0
to 2000 pc cm$^{-3}$, with spacing determined by the
dispersion smearing across each individual frequency channel.
Periodicities were searched for using both a Fast Fourier Transform
and a Fast Folding Algorithm. Known interference frequencies were
eliminated and new pulsar candidates were identified through
inspection of diagnostic plots. A single-pulse search was also
performed; due to the large amount of impulsive interference in our
data, the results of this search are still under analysis.

\section{Results}

Out of 300 fields observed so far, we have processed 214 fields, 
covering about 100 square degrees of sky and redetected 11 known pulsars. 
Three new pulsars, PSRs J0026+6320, J2208+5500 and J2218+5729, have been 
discovered so far. The discovery plots of the new pulsars alongwith 
their average profiles observed at the GMRT and Lovell Telescope are 
shown in Figure 1. The observed parameters of the new pulsars are given in 
Table 1.


The entire data is being reprocessed with better radio frequency 
interference excision 
to look for sources similar to recently reported Rotating Radio 
Transient (McLaughlin et al. 2006), for which the parameters 
of this survey are particularly suitable. 
We are also extending the survey area 
to (45$^\circ$ < l  < 165$^\circ$ ; |b| < 3$^\circ$) and plan 
to complete these observations in the coming months.

\bibliographystyle{aipproc}

\end{document}